\documentclass[%
reprint,
superscriptaddress,
amsmath,amssymb,
aps,
]{revtex4-2}

\usepackage{graphicx}
\usepackage{dcolumn}
\usepackage{bm}
\usepackage[dvipsnames,svgnames]{xcolor}
\usepackage{hyperref}
\hypersetup{
        colorlinks=true,
        citecolor=SteelBlue,
        filecolor=LimeGreen,
        linkcolor=SlateBlue,
        urlcolor=DarkOrchid
}

\begin{document}


\title{Vacuum Beam Guide for Large Scale Quantum Networks}

\author{Yuexun Huang}
\email{yesunhuang@uchicago.edu}
\affiliation{Pritzker School of Molecular Engineering, University of Chicago, Chicago, Illinois 60637, USA}
\author{Francisco Salces\,--\,Carcoba}
\author{Rana X Adhikari}%
\affiliation{Division of Physics, Math, and Astronomy, LIGO Laboratory, California Institute of Technology, Pasadena, CA 91125 USA}
\author{Amir H. Safavi-Naeini}
\affiliation{Department of Applied Physics and Ginzton Laboratory,
Stanford University, Stanford, CA 94305 USA}
\author{Liang Jiang}
\email{liang.jiang@uchicago.edu}
\affiliation{Pritzker School of Molecular Engineering, University of Chicago, Chicago, Illinois 60637, USA}





\date{\today}

\begin{abstract}
The vacuum beam guide (VBG) presents a completely different solution for quantum channels to overcome the limitations of existing fiber and satellite technologies for long-distance quantum communication. With an array of aligned lenses spaced kilometers apart, the VBG offers ultra-high transparency over a wide range of optical wavelengths. With realistic parameters, the VBG can outperform the best fiber by three orders of magnitude in terms of attenuation rate. Consequently, the VBG can enable long-range quantum communication over thousands of kilometers
with quantum channel capacity beyond $10^{13}$ qubit/sec,
orders of magnitude higher than the state-of-the-art quantum satellite communication rate. Remarkably, without relying on quantum repeaters, the VBG can provide a ground-based, low-loss, high-bandwidth quantum channel that enables novel distributed quantum information applications for computing, communication, and sensing.
\end{abstract}

\maketitle


It is an outstanding challenge to build an effective low-loss quantum channel for global-scale quantum networks, which will enable transformative applications of secure quantum communication \cite{XuF20}, distributed quantum sensing \cite{Gottesman12}, and network-based quantum computation \cite{kimble2008quantum, wehner2018quantum}. The main obstacle is the absorption loss of optical channels, with attenuation length limited to tens of kilometers for fiber and free-space channels, resulting in an exponential decrease in the direct quantum communication rate over long distances.
Significant progress has been made in extending the communication distances for quantum networks, including satellite-based quantum entanglement distribution over 1200\,km~\cite{ YinJ17}  and memory-enhanced quantum communication \cite{Bhaskar20} beyond the repeater-less bounds \cite{TGW, PLOB}. However, satellite-based quantum channels are fragile, expensive, and restricted by local weather conditions \cite{Liao17,Goswami23}. Furthermore, quantum repeaters without full quantum error correction will still suffer from a polynomial decrease in communication rate over long distances \cite{Muralidharan16b}.
Hence, it is highly desirable to establish a reliable quantum channel capable of directly transmitting quantum signals over vast distances, such as the continental scale of $10^4$ km (with an attenuation rate of at the level of $10^{-4}$ dB/km) for a wide range of optical frequencies. 

To overcome this challenge, we explore a completely different approach -- the vacuum beam guide (VBG) –- which uses an array of lenses (in an evacuated tube) to guide light, as opposed to relying on total reflection induced by fiber. The large vacuum spacing between lenses significantly reduces the effective travel length of light in optical materials, thus eliminating the problem of material absorption. Inspired by quantum communication satellites, the VBG channel is set up within a vacuum chamber tube, which eliminates air absorption and effectively isolates the channel from the outer environment, ensuring robustness against environmental perturbations. 
Although beam waveguides were proposed for classical optical communication \cite{goubau1968beam}, it was taken over by low-cost optical fiber, which is sufficient for classical communication, despite its intrinsic loss. For quantum communication, however, it is crucial that the design of the vacuum beam guide has the potential to achieve ultra-low-loss long-distance communication. With the ability to build large-scale vacuum chambers hosting precision optical elements separated by multiple kilometers, we can make exciting scientific discoveries, such as gravitational wave detection by the Laser Interferometer Gravitational-wave Observatory (LIGO)~\cite{Abbott16}. 

\begin{figure*}[tb]
\includegraphics[width=\linewidth]{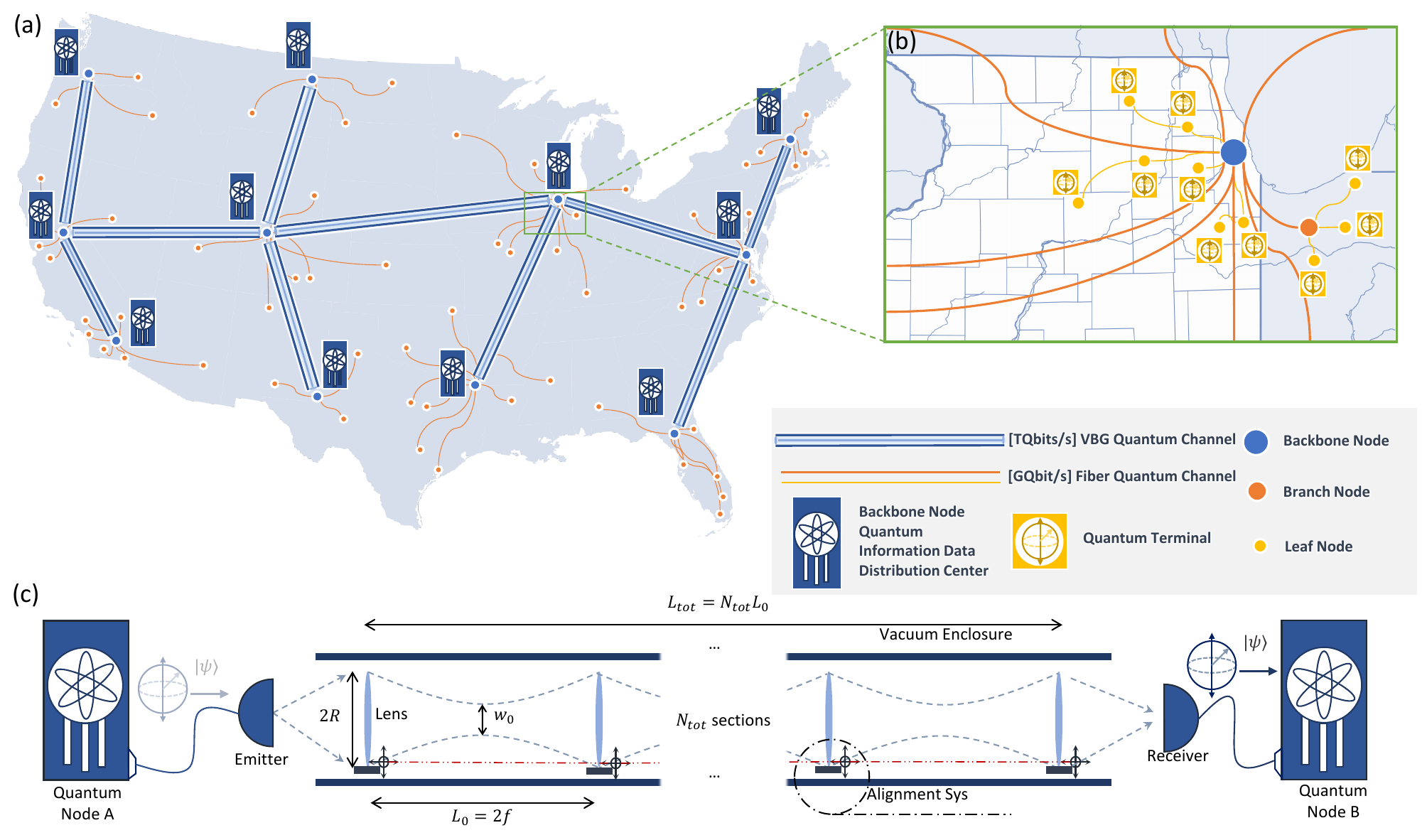}
\caption{\label{fig:BeamGuideIllustration} 
A hierarchical structure of large-scale quantum networks connected by vacuum beam guides (VBGs). 
(a) The VBG backbone network (blue links) can transfer quantum information with high capacity (with TeraQubits/sec) over long distances, connecting regional quantum networks (orange links).
(b) The regional network consists of fiber or free-space quantum channels designed to distribute quantum information to branch nodes and quantum terminals (with GigaQubits/sec) across urban scales.
(c) The design of the VBG involves placing lenses with identical focus length $f$ and radius $R$ at regular intervals within a vacuum chamber tube, constructing a VBG with $N_{tot}$ sections. The positions of the lenses are stabilized by an advanced alignment system. In the VBG, the fundamental Gaussian mode with waist $w_0$ can travel at an extremely low loss over a distance of $L_{tot}$. Quantum states can be transferred from quantum node A at one end of the VBG to quantum node B at the other end of the VBG.}
\end{figure*}

Here, we propose deploying the VBG as the backbone quantum channel toward a global quantum network with a hierarchy structure as shown in Fig.~\ref{fig:BeamGuideIllustration}(a-b). We model the system and estimate the upper bound of the attenuation dominated by residue air absorption, optical losses introduced by lenses, and misalignment of the beam guide. Our estimation demonstrates the VBG as state-of-the-art under a practical and exemplifying configuration, establishing it as one of the most practical and potentially useful quantum communication techniques at a global scale.

\begin{figure*}[tb]
\includegraphics[width=\linewidth]{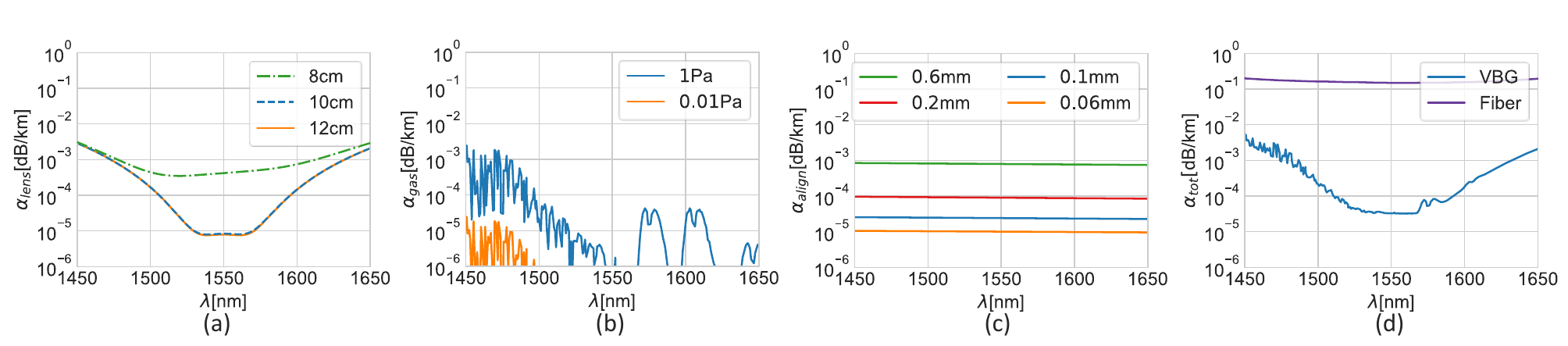}
\caption{\label{fig:Attenuation}Different effective attenuation rates as a function of wavelength under various configurations. (a) Attenuation rate from the lenses, $\alpha_{\mathrm{lens}}$, with lens radius $R=8,\,10,\,20$ cm. For a large radius ($R \ge 10$\,cm), the lens loss is limited by residual reflection and absorption from the AR coating. 
(b)Attenuation rate due to residual gas, $\alpha_{\mathrm{gas}}$, with moderate vacuum pressures: $P=0.01, 1$ Pa. 
(c) The misalignment attenuation rate, $\alpha_{\mathrm{align}}$, with random transverse displacements $\sigma_s=0.06,\,0.1,\,0.2,\,0.6$\,mm while fixing $\frac{\sigma_{L_0}}{L_0}=\frac{\sigma_f}{f}=0.1\%$. (d) The total attenuation rate of the VBG compared to the advanced fiber \cite{nagayama2002ultra,sakr2021hollow} 
under the configuration of 
$L_0 = 4$\,km, $R=10$\,cm, $P=1$\,Pascal, and $\sigma_s=0.1$\,mm.}
\end{figure*}

\paragraph*{Ultra-High Transmission of VBG.}
The VBG is a long vacuum chamber tube that consists of an array of $N_{\mathrm{tot}}$ lenses spaced $L_{0}$ apart, which enables the connection of quantum terminals separated by $L_{\mathrm{tot}}= N_{\mathrm{tot}} L_{0}$, as illustrated in Fig.~\ref{fig:BeamGuideIllustration}(c). 
The vacuum has a typical pressure below $\sim1$ Pascal, which ensures low absorption from the remaining gas at room temperature. The lenses are shielded 
from seismic vibrations and are optically aligned with adaptive feedback. In this analysis, we consider a feasible and robust confocal design with a spacing of $L_{0}=4$\,km and a focal length of $f= L_{0} /2$. The beam waist is $w_0 = \sqrt{\lambda f / \pi} \approx 3$ cm for the telecom-band wavelength $\lambda$. The lens radius of $R=10$\,cm is sufficient to achieve negligible diffraction loss. 

Optical signals encoding quantum information can transmit through the VBG with little loss, even after passing through an array of aligned lenses over thousands of kilometers. The effective attenuation rate (in units of dB/km) characterizes the VBG transmission loss, with three major contributions associated with the lens, residual gas, and imperfect alignment: 
\begin{equation}
\label{eq:totalLoss}
\alpha_{\mathrm{tot}}=\alpha_{\mathrm{lens}}+\alpha_{\mathrm{gas}}+\alpha_{\mathrm{align}}.
\end{equation}
We now discuss the conditions to achieve an attenuation rate at the level of $10^{-4}$ dB/km for all three contributions.

The lens loss, $\alpha_{\mathrm{lens}}$, is associated with absorption, scattering, reflection, and diffraction. 
As illustrated in Fig.~\ref{fig:Attenuation}(a), a lens radius of $R \ge 10$ cm is adequate to suppress diffraction loss for $L_{0}=4$ km. 
Furthermore, by adding a multi-layer anti-reflective coating, the total loss per lens can be reduced to less than $10^{-4}$, resulting in an effective loss rate of $\alpha_{\mathrm{lens}} < 10^{-4}$ dB/km at the wavelengths $\lambda \approx 1550$ nm, which can match the telecom band \cite{SM}.

Gas loss in the VBG is primarily due to the absorption from residual air in the vacuum chamber tube. As shown in Fig.~\ref{fig:Attenuation}(b), we compute the attenuation rate $\alpha_{\mathrm{gas}}$ at various levels of gas pressure with the components of air based on the HITRAN database~\cite{kochanov2016hitran}. At a pressure of 1 Pascal, the attenuation rate is $\alpha_{\mathrm{gas}} < 10^{-4}$ dB/km for optical wavelengths within the selected telecom bands. Further reducing the air pressure below $10^{-2}$ Pascal is sufficient to achieve negligible air absorption by reducing attenuation rate below $10^{-4}$ dB/km almost over the entire spectrum. 


We can derive an upper bound for the effective attenuation rate induced by imperfect alignment in the confocal design:
\begin{equation}
\label{eq:perturbationLoss}
\alpha_{\mathrm{align}}\leq -\frac{10}{L_{0}} \log_{10} \left[1-\frac{2\sigma_s^2}{w_0^2}
-\frac{\sigma_{L_{0}}^2}{L_0^2}
-\frac{\sigma_f^2}{f^2} \right], 
\end{equation}
where $\sigma_s$ and $\sigma_{L_{0}}$ are the magnitudes of the fluctuations of transverse and longitudinal displacements for each lens, and $\sigma_f$ is the deviation of the focal length. The general bound for $\alpha_{\mathrm{align}}$ for near confocal design is derived in Supplementary Material \cite{SM}.
Since $w_0 \ll L_{0},\,f$, the fluctuating transverse displacement is the dominant cause of attenuation, while the other two contributions can be sufficiently small, $\frac{\sigma_{L_{0}}}{L_0}, \frac{\sigma_f}{f} < 10^{-3}$, achievable with current technology 
\cite{SM}.
As illustrated in Fig.~\ref{fig:Attenuation}(c), it is essential to have good transverse alignment (e.g., $\sigma_s < 0.2$ mm) to achieve a low effective attenuation rate ($\alpha_{\mathrm{align}} < 10^{-4}$ dB/km). Even with relatively poor alignment (e.g., $\sigma_s = 0.6$ mm), we can still achieve $\alpha_{\mathrm{align}} < 10^{-2}$ dB/km, which is much better than the attenuation rate of fiber by at least an order of magnitude. Practically, each lens will be actively positioned using slow actuators, based on standard alignment sensing systems~\cite{Dooley:13}, bringing the residual mis-centering to $< 0.1$\,mm.

By summing up all three contributions, we can plot the total attenuation rate along with the individual contributions in Fig.~\ref{fig:Attenuation}(d), assuming practical parameters such as $R=10$ cm, $P=1$ Pascal, and $\sigma_s = 0.1$ mm. The VBG can achieve an attenuation level as low as $5\times10^{-5}$ dB/km for an optimized choice of wavelength within the atmospheric window, which corresponds to an effective attenuation length of 80,000\,km, more than three orders of magnitude better than state-of-the-art fiber technology.

\paragraph*{Quantum Channel Capacities of VBG.}
We can use the VBG as a highly transparent bosonic quantum channel to transmit quantum information over long distances \cite{holevo2001evaluating}. 
The VBG has a transmission efficiency at wavelength $\lambda$
\begin{equation}
    \begin{aligned}
    \eta[\lambda] &=10^{-0.1 L_{\mathrm{tot}} \alpha[\lambda]}, \\
    \end{aligned}
\end{equation}
which can be used for various quantum communication protocols.
For one-way quantum communication (e.g., from Alice to Bob only), the one-way pure-loss capacity for a single wave packet at a wavelength $\lambda$ is given by
\begin{equation}
    \begin{aligned}
    q_1[\lambda] &=\max \left[\log_2\left(\frac{\eta[\lambda]}{1-\eta[\lambda]}\right),0\right], \\
    \end{aligned}
\end{equation}
which vanishes for $\eta[\lambda]\le 1/2$.
If the pure-loss VBG channel is further assisted by two-way classical communication (between Alice and Bob), the corresponding two-way pure-loss capacity at a wavelength $\lambda$ is \cite{pirandola2017fundamental}
\begin{equation}
    \begin{aligned}
    q_2[\lambda] &=\log_2\left(\frac{1}{1-\eta[\lambda]}\right). 
    \end{aligned}
\end{equation}
which is finite for all $\eta[\lambda]>0$.

\begin{figure*}[t]
\includegraphics[width=\linewidth]{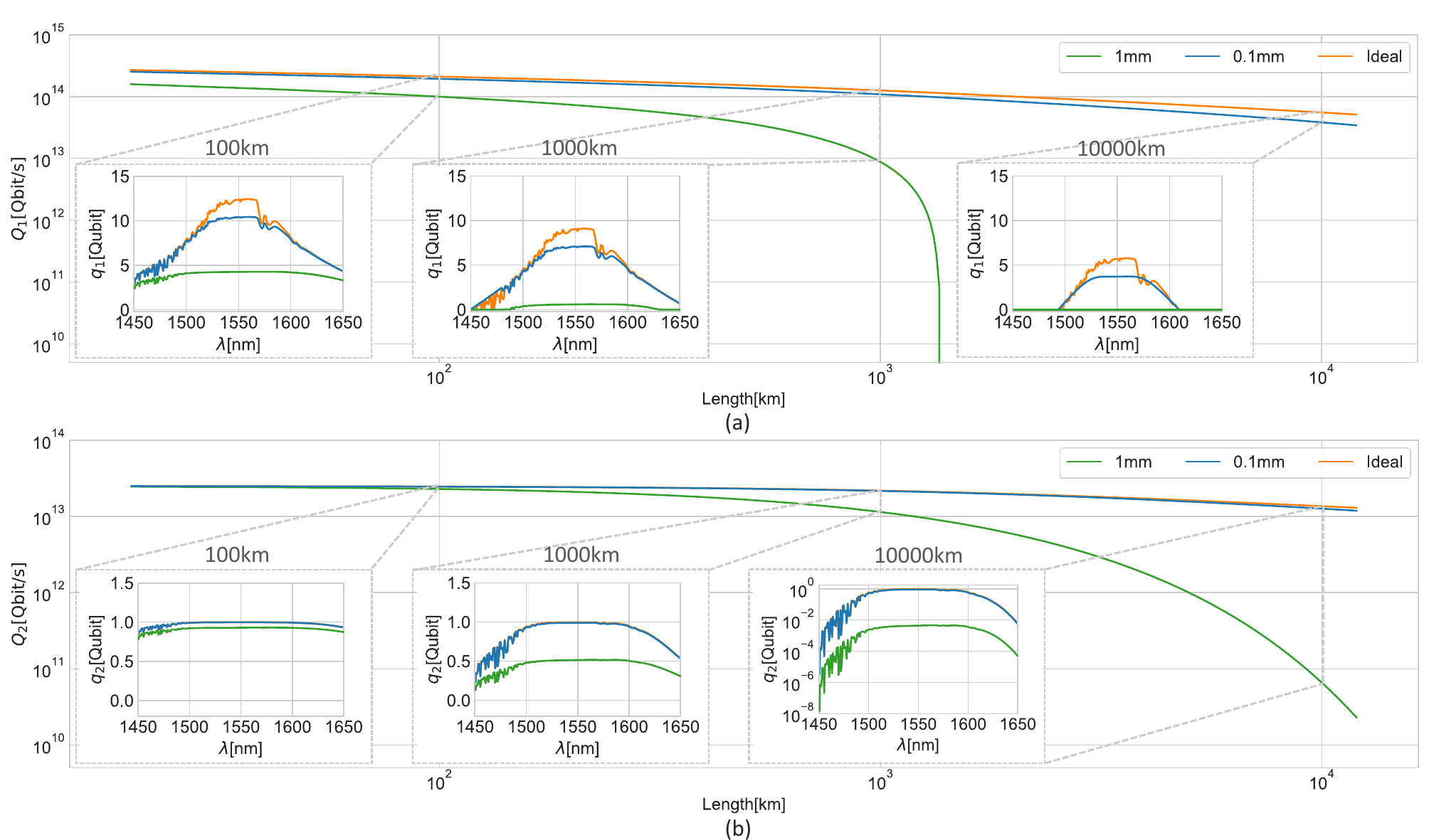}
\caption{\label{fig:ChannelCapacity} The quantum channel capacity $Q_{1(or2)}$ at $1$ Pascal as a function of transmission distance with $\sigma_s=0.1,\,1$ mm when fixing $\frac{\sigma_{L_0}}{L_0}=\frac{\sigma_f}{f}=0.1\%$ compared to the ideal alignment. (a) One-way frequency-integrated quantum capacity with perfect coupling. (b) Two-way frequency-integrated quantum capacity assuming $50\%$ coupling efficiency. The insets show the relation between channel capacity $q_{1 (\mathrm{or} 2)}$ as a function of wavelength at $L = 100,\,1000,\,10000$ km. The inset of $10000$ km in (b) is in log scale to show finite two-way quantum capacity over a wide range of parameters. 
}
\end{figure*}

As shown in the insets of Fig.~\ref{fig:ChannelCapacity}, the VBG can have a broad band with a large one-way or two-way channel capacity over long distances. In the low-loss regime with $L_{\mathrm{tot}} \ll 1/\alpha$, efficient multi-mode encoding techniques can be used to approach the asymptotic scaling of one-way channel capacity~\cite{Noh18}.

We can calculate the frequency-integrated channel capacity, which provides the maximum transmission rate of quantum information, by integrating over the frequency~\cite{wang2022quantum}: 
\begin{equation}
Q_{1 (\mathrm{or} 2)} \equiv \int q_{1 (\mathrm{or} 2)} d\nu = \int q_{1 (\mathrm{or} 2)}[\lambda] \frac{c}{\lambda^2} d\lambda,
\end{equation}
for one-way and two-way quantum communication protocols, respectively.
The VBG has a significant advantage over other techniques, particularly due to its ultra-low attenuation rate over a wide range of wavelengths, leading to an extremely large quantum capacity exceeding $10^{13}$ qubits/second over a distance of $10^4$\,km under one-way quantum communication protocol as shown in Fig.~\ref{fig:ChannelCapacity}(a). While the estimation in Fig.~\ref{fig:ChannelCapacity}(b) suggests that even with a baseline coupling loss as large as $50\%$, where the one-way protocol fails, it is still possible to achieve a Tera-level qubit rate via two-way protocols for continental scale communication. This points out a practically feasible pathway toward ultra-fast global quantum networks.

Compared to state-of-the-art satellite-ground links for quantum communication \cite{ChenYA21}, the ground-based VBG offers high reliability as it is operational at all times, and provides an extremely high throughput, at least eight orders of magnitude higher in terms of achievable quantum communication rate. 
The concept of a beam guide \cite{goubau1968beam} was investigated to address diffraction loss in satellite-relayed quantum communication \cite{Goswami23}. Nevertheless, that protocol remains susceptible to atmospheric and turbulence-induced losses, thereby affecting reliability, transmission rate, channel capacity, and communication latency (due to the limitation of two-way quantum communication).
Additionally, unlike quantum repeaters \cite{Azuma22}, the VBG only requires passive optical lenses to focus the optical beams and does not need any quantum memory or active quantum devices to perform quantum error detection or correction. Therefore, the VBG should be feasible to implement with current technology.

\paragraph*{Discussion.}
We can also design the VBG to operate with visible light, which would benefit from a larger air transmission window and low-loss lens materials if sufficiently small lens roughness is achievable, 
while better transversal alignment is required to match the reduced beam waist $w_0 \propto \lambda ^{1/2}$ with shorter wavelengths. In practice, there will be a trade-off between the VBG's performance and cost, and we can optimize its design parameters to achieve the desired balance.


\paragraph*{Summary and Outlook.}
We have presented a ground-based VBG scheme that enables the implementation of a highly transparent and reliable optical quantum channel with an effective attenuation length of over $10^4$ km and a large communication bandwidth. With currently available technology, the VBG can establish a continuous quantum channel connecting remote quantum devices with an ultra-high quantum capacity above $10^{13}$ qubits/second over continental scales, orders of magnitude higher than other approaches using satellites and quantum repeaters. 

By addressing the challenge of lossy quantum channels, our high-throughput VBG has the potential to revolutionize quantum networks, enabling a wide range of exciting novel quantum network applications, such as global-scale secure quantum communication \cite{XuF20}, ultra-long-based optical telescopes \cite{Gottesman12}, quantum network of clocks \cite{Komar14}, quantum data centers \cite{LiuJ22b}, and delegated quantum computing \cite{barz2012demonstration,Fitzsimons17}.

\paragraph*{Acknowledgments}
We thank David Awschalom, Saikat Guha, Dan Brown,  Ming Lai, John Preskill and Peter Zoller for helpful comments and discussions.
We acknowledge support from the ARO(W911NF-23-1-0077), ARO MURI (W911NF-21-1-0325), AFOSR MURI (FA9550-19-1-0399, FA9550-21-1-0209, FA9550-23-1-0338), NSF (PHY-0823459, PHY-1764464, ECCS-1941826, OMA-1936118, ERC-1941583, OMA-2137642, OSI-2326767, CCF-2312755), NTT Research, Packard Foundation (2020-71479), and the Marshall and Arlene Bennett Family Research Program. This material is based upon work supported by the U.S. Department of Energy, Office of Science, National Quantum Information Science Research Centers.

\bibliography{VacuumBeamGuide}

\newpage
\clearpage



\end{document}